\documentclass[lettersize,journal]{IEEEtran}
\usepackage[square,comma]{natbib}
\setcitestyle{numbers,square}
\usepackage{amsmath,amsfonts}
\usepackage{hyperref}
\usepackage[capitalize]{cleveref}
\crefname{section}{section}{sections}
\usepackage{url}            % simple URL typesetting
\usepackage{algorithmic}
\usepackage{algorithm}
\usepackage{array}
\usepackage{xcolor}
\usepackage{graphicx}
\usepackage{subcaption}
\usepackage{textcomp}
\usepackage{stfloats}
\usepackage{verbatim}
\usepackage{enumitem}
\usepackage{multirow}
\usepackage{colortbl}

\definecolor{red}{rgb}{1,0,0}
\definecolor{headerColor}{RGB}{216, 216, 216}

\usepackage{caption}
\newcommand{\revision}[1]{{\color{black}#1}}
\newcommand{\revisionsec}[1]{{\color{black}#1}}
\newcommand{\revisionminor}[1]{{\color{black}#1}}

\setlist[enumerate]{noitemsep, topsep=0pt, leftmargin=*}

\DeclareGraphicsExtensions{.pdf,.jpg,.png}

\newcommand{\para}[1]{\vspace{0.05in}\noindent{\bf #1}\quad}

\begin{document}
	
\title{Towards Open-Vocabulary Video Semantic Segmentation}

% \author{
    %   \IEEEauthorblockN{Author Name 1\IEEEauthorrefmark{1}, Author Name 2\IEEEauthorrefmark{2}, and Author Name 3\IEEEauthorrefmark{3}}
    %   \IEEEauthorblockA{\IEEEauthorrefmark{1}Department of Electrical Engineering, University Name, City, Country \\
        %     Email: author1@university.edu}
    %   \IEEEauthorblockA{\IEEEauthorrefmark{2}Department of Computer Science, University Name, City, Country \\
        %     Email: author2@university.edu}
    %   \IEEEauthorblockA{\IEEEauthorrefmark{3}Department of Mechanical Engineering, University Name, City, Country \\
        %     Email: author3@university.edu}
    %   \IEEEauthorblockN{IEEE Publication Technology,~\IEEEmembership{Staff,~IEEE,}}
    % }

\author{Xinhao Li, Yun Liu, Guolei Sun, Min Wu, Le Zhang\thanks{The corresponding author is Le Zhang.} and Ce Zhu,~\IEEEmembership{Fellow,~IEEE}
    % <-this % stops a space
    \thanks{Xinhao Li, Le Zhang and Ce Zhu are with the School of Information and Communication Engineering, University of Electronic Science and Technology of China (UESTC), Chengdu 611731, China.}%
    \thanks{Yun Liu and Guolei Sun are with the College of Computer Science, Nankai University, Tianjin 300350, China.}%
    % \thanks{Guolei Sun is with Computer Vision Lab, ETH Zurich, Zurich, Switzerland.}%
    \thanks{Min Wu is with the Institute for Infocomm Research (I2R), Agency for Science, Technology, and Research (A*STAR), Singapore.}% <-this % stops a space.
}



\maketitle

\begin{abstract}
\revision{Semantic segmentation in videos has been a focal point of recent research. However, existing models encounter challenges when faced with unfamiliar categories. To address this, we introduce the Open Vocabulary Video Semantic Segmentation (OV-VSS) task, designed to accurately segment every pixel across a wide range of open-vocabulary categories, including those that are novel or previously unexplored. To enhance OV-VSS performance, we propose a robust baseline, OV2VSS, which integrates a spatial-temporal fusion module, allowing the model to utilize temporal relationships across consecutive frames. Additionally, we incorporate a random frame enhancement module, broadening the model’s understanding of semantic context throughout the entire video sequence. Our approach also includes video text encoding, which strengthens the model's capability to interpret textual information within the video context.
Comprehensive evaluations on benchmark datasets such as VSPW and Cityscapes highlight OV-VSS’s zero-shot generalization capabilities, especially in handling novel categories. The results validate OV2VSS's effectiveness, demonstrating improved performance in semantic segmentation tasks across diverse video datasets. The source code will be made available at: \href{https://github.com/AVC2-UESTC/OV2VSS}{https://github.com/AVC2-UESTC/OV2VSS}.}
\end{abstract}


\section{Introduction}
% \IEEEPARstart{T}{his} file is intended to serve as a ``sample article file''
\IEEEPARstart{S}{emantic} \revisionsec{segmentation, the task of classifying each pixel into distinct semantic categories \cite{jin2021mining,jin2021isnet,zhou2024boundary,gao2022fbsnet,yan2019semantic,gou2024difference}, is a core component of computer vision with substantial practical value. However, traditional segmentation models are constrained by being trained on a fixed set of predefined categories. Consequently, they can only recognize categories present in the training data, presenting a significant challenge for real-world applications where novel or unseen categories frequently emerge. This limitation hinders the model's adaptability and scalability in dynamic, diverse environments.}

\begin{figure*}[!t]
    \centering
    \includegraphics[width=.9\linewidth]{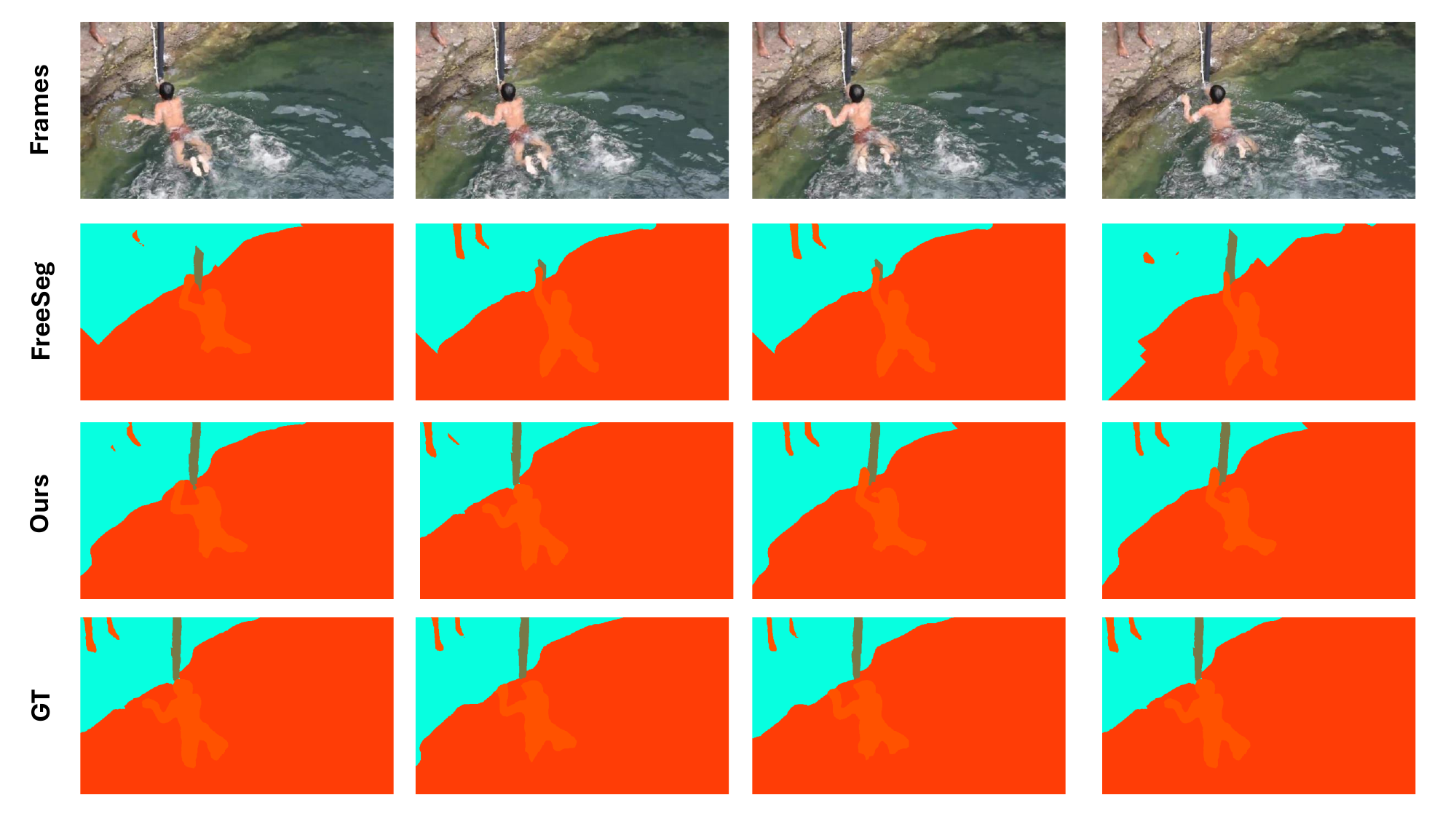}
    \vspace{-.10in}
    \caption{The comparison between our method and image-based methods on the VSPW dataset, from top to bottom, includes the original frame, image method, our method, and ground truth. From the figure, it can be seen that image-based methods are prone to producing discontinuities in the image.}
    \label{fig:compare}
    \vspace{-.15in}
\end{figure*}

\revision{Recent progress has been made in addressing this issue, particularly in image-based semantic segmentation. Notably, studies have integrated the CLIP model~\cite{radford2021learning} into semantic segmentation \cite{xu2023side,ghiasi2022scaling} to achieve open-vocabulary image semantic segmentation. These efforts leverage vision-language knowledge to enhance generalization capabilities, particularly for unseen categories. During testing, the model is capable of segmenting unseen categories as long as their names are provided.

In contrast, the challenge of open-vocabulary video semantic segmentation (\textbf{OV-VSS}) has not been systematically explored, despite the fact that our real-world experiences are better represented by dynamic video content rather than static images. Video-based semantic segmentation presents additional complexities due to the temporal dimension, which captures evolving scenes, objects, and contextual information that cannot be encapsulated in a single image. Studying VSS is essential because videos are more prevalent in practical applications, such as autonomous driving, surveillance, and activity recognition, where recognizing objects and actions over time is critical.

While a straightforward solution might be to apply existing image-based methods, such as SAN \cite{xu2023side}, to video datasets, this approach neglects the rich temporal context inherent to video footage. To validate this hypothesis, we conducted a series of experiments by applying image-based methods to video datasets. The results, as shown in \cref{fig:compare}, reveal significant performance degradation, emphasizing the need for a more refined approach to VSS. These findings will be further elaborated upon in the subsequent sections.}

\begin{figure}[!t]
    \centering
    \includegraphics[width=\linewidth]{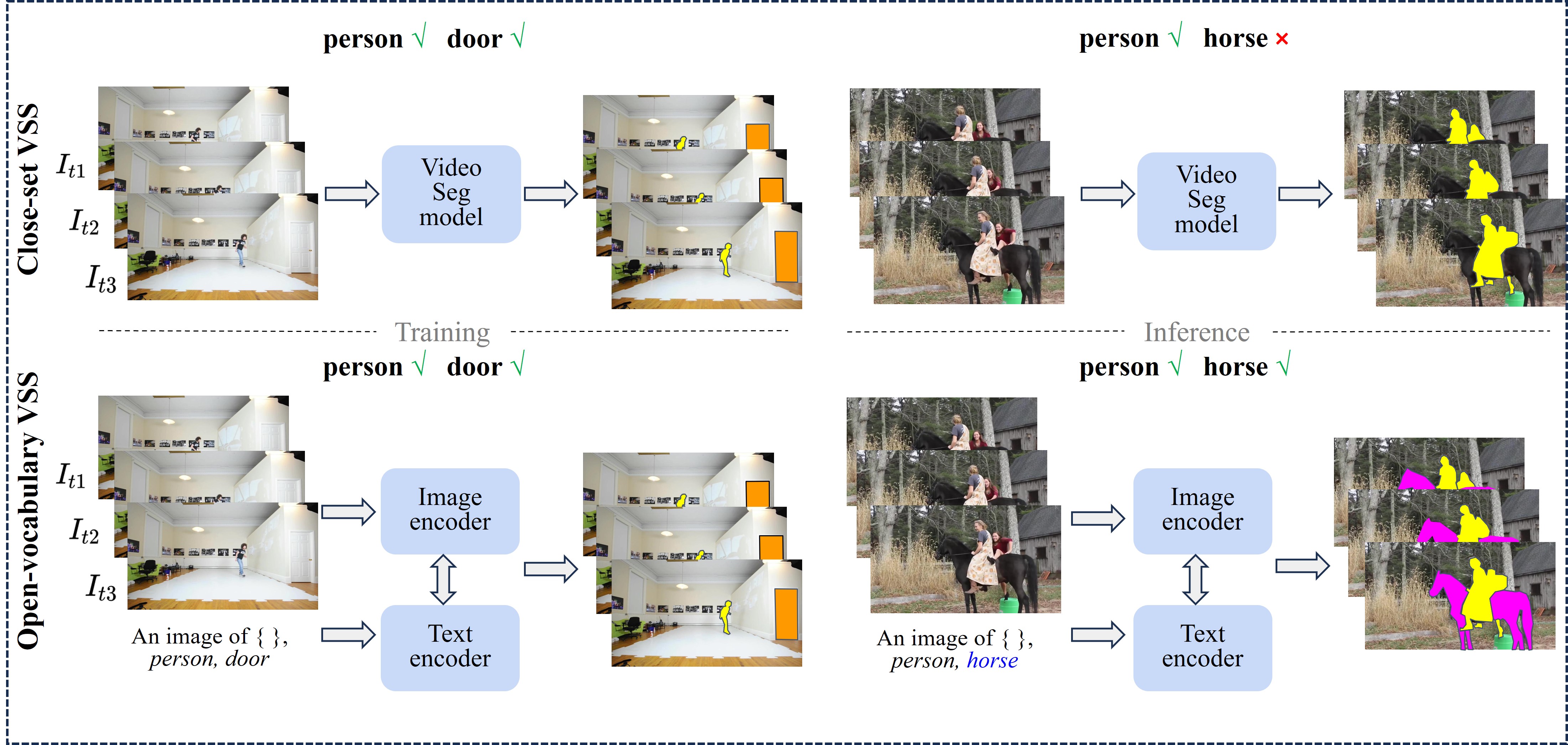}
    \vspace{-.10in}
    \caption{Comparison of VSS and OV-VSS: In traditional VSS, the model is trained on a closed set of classes (e.g., "person" and "door") and fails to segment novel classes (e.g., "horse," as shown in the figure)(upper). In contrast, our Open-Vocabulary Segmentation model is trained on base classes but can simultaneously segment both base and novel categories(lower).}
    \label{fig:intro}
    \vspace{-.15in}
\end{figure}

% Concretely, given a training dataset $\mathcal{D}_{\text{train}}$ containing pixel-level annotations for each training-seen category, a traditional model $f_\theta(\cdot)$ aims to train a model that aims to segment each class in $\mathcal{D}_{\text{test}}$, while novel categories are ignored. In contrast, Open-vocabulary Video Semantic Segmentation (Open-VSS) aims to train a model $f_\theta(\cdot)$ on $\mathcal{D}_{\text{train}}$, and then test on $\mathcal{D}_{\text{test}}$ for both training categories $\mathcal{C}_{\text{train}}$ and novel categories $\mathcal{C}_{\text{novel}}$. Specifically, given a test video $V_i \in \mathbb{R}^{T_i \times H_i \times W_i \times 3}$ during test, the trained model $f_\theta(\cdot)$ is supposed to segment each category mask and category label. The full process is shown in \cref{fig:intro}.

 \revision{Inspired by this, we introduce the challenge of open-vocabulary video semantic segmentation (OV-VSS), which differs from conventional VSS in its ability to segment and classify objects not predefined during training. The key distinctions are illustrated in \cref{fig:intro}.   In OV-VSS, our goal is to classify every pixel within each video frame, including categories the model has not previously encountered. To address this task, we propose a robust baseline model, OV2VSS, which integrates a spatial-temporal fusion module that leverages temporal relationships across consecutive frames. Additionally, we include a random frame enhancement module to expand the model’s understanding of semantic context throughout the entire video sequence. Our approach also incorporates video text encoding to enhance the model’s ability to interpret textual information within the video, enriching its semantic comprehension.

 It is worth noting that while some recent attempts have addressed open-vocabulary segmentation in videos, these methods \cite{wang2023openvocabulary, guo2023openvis} focus solely on instance segmentation. In contrast, our open-vocabulary semantic segmentation (VSS) task provides a broader and more comprehensive approach by classifying every pixel within each frame across a wide range of categories, including both foreground objects and the surrounding context. This distinction is crucial because our method goes beyond isolated instances to deliver a full scene understanding, capturing both object and non-object elements, such as ``road",``sky", or ``vegetation", which may lack clear instance boundaries.

The capability of VSS to label both primary objects and contextual regions enriches video content representation, supporting applications where an in-depth understanding of the entire scene is essential. For instance, autonomous vehicles benefit from VSS’s ability to identify drivable areas, sidewalks, and other critical scene elements that may not constitute distinct object instances but are vital for navigation. Similarly, in augmented reality, where overlaying information across entire scenes is important, VSS provides a more immersive and accurate experience by segmenting and labeling each part of the environment.

  Methodologically, open-vocabulary instance segmentation (OV-VIS) typically employs CLIP as a classifier to map object queries (proposals) to open-vocabulary labels. This approach relies on object-centric queries to propose instances, which are then classified based on text embeddings from CLIP’s text encoder. As a result, OV-VIS performance is constrained by the quality of its query proposal system, with segmentation accuracy heavily depending on how effectively the model can generate and classify object instances.

In contrast, our approach to OV-VSS is fundamentally different. We leverage both CLIP’s image and text encoders to model pixel-level correlations between visual features and text prompts, allowing us to focus on pixel-wise semantic information across the entire video sequence rather than relying on object proposals. This enables us to utilize both spatial and temporal information, ensuring accurate labeling for all pixels—whether belonging to objects or background elements. The inclusion of our Spatio-Temporal Fusion and Random Frame Enhancement modules further strengthens our approach, enabling efficient capture of temporal dependencies between frames, which is critical for consistent segmentation over time.
%感觉这两段可以合并一下
% While OV-VIS is limited by its dependence on object proposals, our OV-VSS approach performs global pixel-wise classification, bypassing the need for explicit object-level segmentation. This makes our method more scalable and better suited for tasks requiring a complete understanding of video scenes, as it captures broader semantic relationships beyond individual objects.
	
In our experiments, we evaluate our methods using the VSPW datasets \cite{miao2021vspw}, comprising 124 classes. For clarity, we categorize the datasets into seen classes and unseen classes, maintaining as 80:44. On the challenging VSPW dataset \cite{miao2021vspw}, we achieve a mIoU of $17.99\%$ on unseen classes, demonstrating a notable $4\%$ improvement compared to the current state-of-the-art in image-based open-vocabulary segmentation. 
}

To sum up, the contributions of this paper are as follows:
	\begin{enumerate}
		\item \revision{We introduce the novel task of Open Vocabulary Video Semantic Segmentation (OV-VSS), demonstrating its significance in addressing the limitations of traditional video semantic segmentation models that struggle with unseen categories.}

        \item \revision{We propose a new baseline method, OV2VSS, which integrates a Spatial-Temporal Context Fusion module to capture intra-frame correlations and enhance temporal consistency. Additionally, we introduce a Random Frame Enhancement module to leverage long-range frame dependencies, alongside a specialized Video-Text Encoding module to align video content with textual descriptions.}

        \item \revision{We conduct extensive evaluations on the VSPW dataset, showcasing the effectiveness and generalizability of our approach. Our method achieves state-of-the-art performance in segmenting unseen classes, significantly improving over previous image-based approaches.}
	\end{enumerate}
	
	\section{Related Work}
	\subsection{Video Semantic Segmentation}
	\revision{Video Semantic Segmentation (VSS) aims to categorize pixels in each video frame according to predefined categories. Early VSS efforts were constrained by datasets like Cityscapes \cite{cordts2016cityscapes} and NYUDv2 \cite{silberman2012indoor}, which provided annotations for only a few sporadic frames. Similarly, the CamVid dataset \cite{brostow2008segmentation} suffers from limited scale and low frame rate. The introduction of the VSPW dataset \cite{miao2021vspw}, with 124 classes and a wide range of scenarios, significantly advanced the field. It features an average of over 70 annotated frames per video, providing rich data for training and evaluation.

On this challenging dataset, existing methods can be divided into two main categories. One category focuses on improving performance with minimal additional computation \cite{li2018low,shelhamer2016clockwork,mahasseni2017budget,zhu2017deep,xu2018dynamic,jain2019accel,hu2020temporally,liu2020efficient,paul2021local}. These methods often prioritize keyframes, applying high-computation networks to keyframes and using lower-computation methods, such as optical flow, to process non-keyframes and capture temporal correlations.

Conversely, another set of methods \cite{kundu2016feature,li2021video,gadde2017semantic,jin2017video,liu2017surveillance,nilsson2018semantic,zhu2019improving,zhang2022auxadapt,sun2022coarse,sun2022mining,sun2024learning,li2022video,li2023tube,weng2023mask} prioritizes segmentation accuracy by deploying resource-intensive networks on each frame to achieve superior results. However, these methods rely heavily on segmenting classes present in the training datasets, limiting their adaptability to unseen classes—a significant challenge for VSS in real-world applications, where new and previously unencountered classes need accurate identification and segmentation.}
	
\subsection{Open-Vocabulary Segmentation}
	
\revision{Vision-language models (VLMs) represent a significant leap in connecting visual and textual modalities. Pre-trained models like CLIP \cite{radford2021learning} and ALIGN \cite{jia2021scaling} harness large datasets to enable impressive zero-shot object recognition capabilities, boosting performance in tasks such as classification, captioning, and segmentation.

In image segmentation, VLMs excel at zero-shot predictions. For example, Xu et al. \cite{xu2023side} propose a side network for open-vocabulary segmentation, while Ding et al. \cite{ding2022decoupling} decouple segmentation and classification. Ghiasi et al. \cite{ghiasi2022scaling} align text queries with image regions.

However, applying VLMs directly to video data introduces new challenges, as models like CLIP are trained on static images. Efforts like Wang's \cite{wang2023openvocabulary} aim to extend VLMs to video instance segmentation, but further research is needed to address the complexities of dynamic visual content.}
	
	\section{Our Methods}\label{sec:methods}
 For a given video clip, a training sample can be represented as follows:
	\begin{equation}
		([I_{t-k_1}, \cdots, I_{t-k_n}, I_t, I_{random}], G_t),
	\end{equation}
	where we define that
	\begin{itemize}
		\item $I_t \in \mathbb{R}^{H \times W}$ denotes the frame of interest, referred to as the target frame at timestamp $t$, and is accompanied by its associated ground-truth segmentation map $G_t \in \mathbb{R}^{H \times W}$.
		\item $I_{t-k_1}, \cdots, I_{t-k_n}$ represent the $n$ preceding frames, each located $k_1, \cdots, k_n$ frames away from $I_t$ ($k_1 > k_2 > \cdots > k_n$). In our experiments, $k_{n}-k_{n+1}$ is set to 3. 
		\item  $I_{\text{random}}$ signifies a temporally distant frame extracted from the same video, contributing temporal correlations to facilitate the training process. During the training stage, a frame is randomly selected as $I_{\text{random}}$. In the inference stage, $I_{\text{random}}$ will be the most temporally-distant frame in the same video. 
	\end{itemize}
\revision{For each frame \(I_{t_i}\), we first pass it through a backbone network to extract intermediate feature maps \(\{U_{t_i}^l \in \mathbb{R}^{H_l \times W_l \times C_l}\}_{l=1}^L\) across different scales from \(L\) layers, where \(H_l, W_l, C_l\) represent the height, width, and channels of the feature map, respectively. We simplify by omitting the superscript \(l\). Next, the spatial-temporal context fusion module integrates the features across the video sequence, producing the refined feature \(O_t\) for the target frame.

The target frame \(I_t\) and class name text \(T(n)\) for \(n = 1,...,N\) are input into the video-text encoding module to derive language-guided features, which are concatenated with \(U_t\) to produce position-aware features \(\hat{X}\). Additionally, the random frame feature \(D_{\text{random}}\) is enhanced using \(O_t\) via the random frame enhancement module, yielding refined feature \(\hat{O}_t\). Finally, the enhanced feature \(\hat{O}_t\) and language-guided feature \(\hat{X}\) are fed into the decoder. The detailed design is discussed in subsequent sections.
}
	
	\begin{figure*}[!t]
		\centering
            \setlength{\abovecaptionskip}{1.0cm}
		\includegraphics[width=\linewidth]{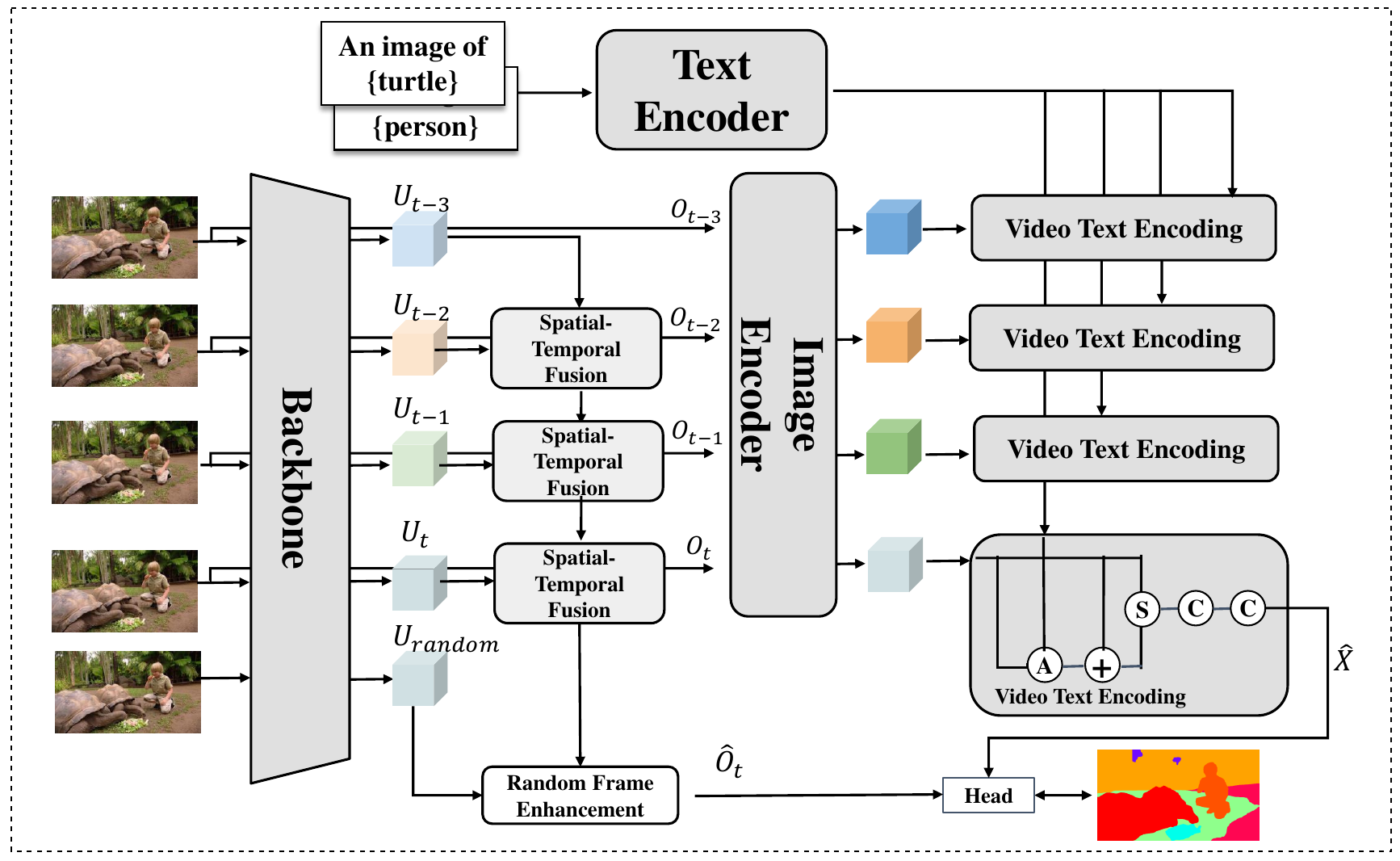}
		\vspace{-12mm}
		\caption{\textbf{Overall structure of OV2VSS}. Our approach primarily leverages three modules: the spatial-temporal information fusion module, which integrates spatial-temporal details from the video; the random frame enhancement module, which acquires contextual information from a randomly selected frame; the video text encoding, which utilizes text supervision in the training process. A denotes attention, + denotes element-wise addition, S denotes establish a cost-volume, C denotes Concatenate. The text encoder and image encoder are obtained from CLIP. Detailed explanations are included in \cref{sec:methods}.}
		\label{fig:pipeline}
	\end{figure*}

	\subsection{Spatial-Temporal Context Fusion}
	\label{sec:stcf}
\revision{The key distinction between video and image semantic segmentation lies in capturing inter-frame temporal relationships. The Spatial-Temporal Context Fusion (STCF) module is specifically designed to integrate temporal information through an information-rich module and a temporal transfer module.

Traditional attention-based temporal fusion methods derive the query from the target frame and use reference frames for keys and values. However, this can cause discontinuities due to differences in categories and environments across frames. As the temporal distance increases, the likelihood of such variations also grows.}

    \begin{figure}[!t]
        \centering
        \includegraphics[width=\linewidth]{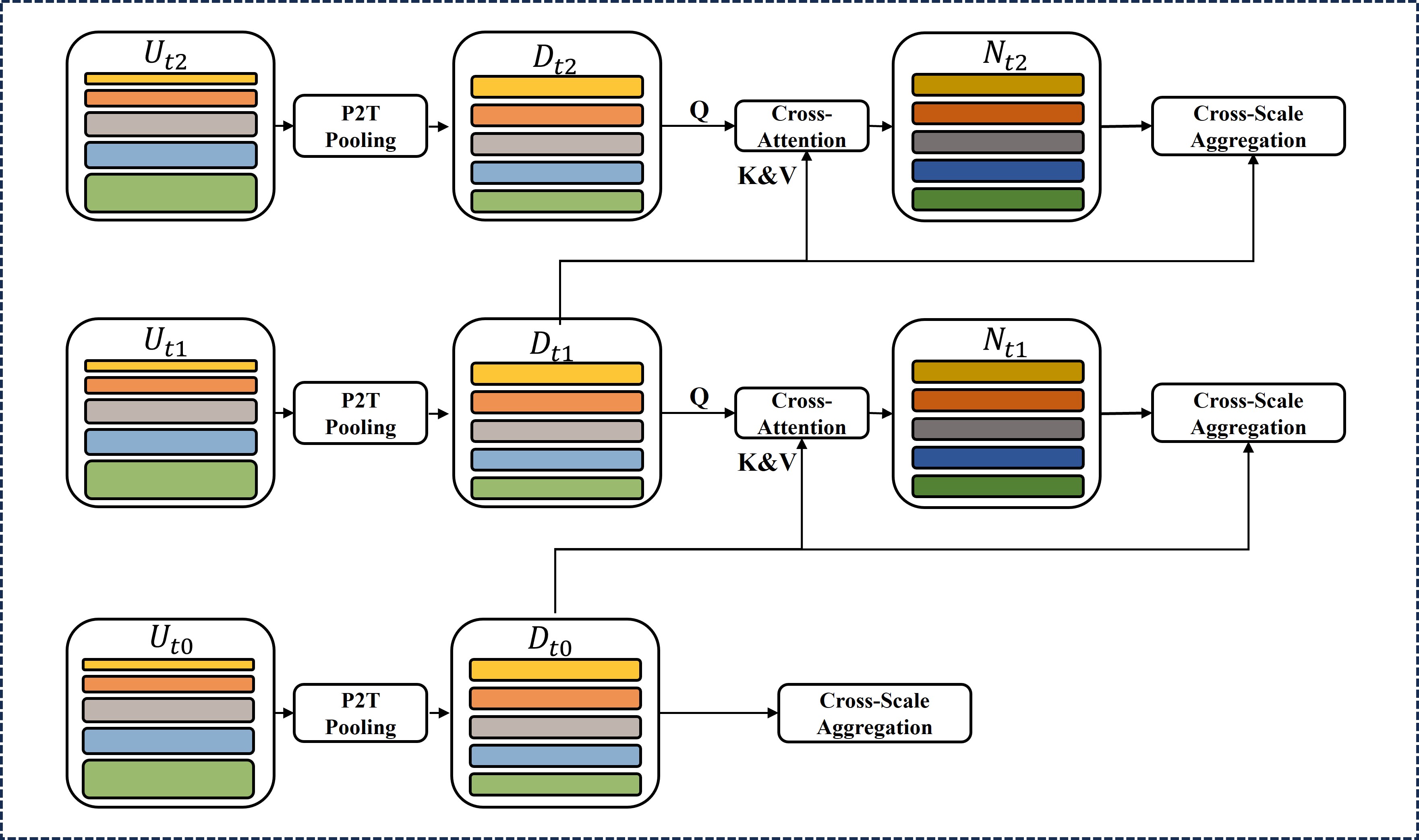}
        \vspace{-.10in}
        \caption{The architecture of the Spatio-Temporal Context Fusion module, which includes a P2T pooling, cross-attention and a cross-scale Aggregation.}
        \label{fig:cross-frame}
        \vspace{-.15in}
    \end{figure}
	
\revision{This insight motivates a gradual aggregation approach: instead of directly fusing \(I_{t-k_1}\) with \(I_t\), we first aggregate the closest frames, such as \(I_{t-k_1}\) with \(I_{t-k_2}\), then \(I_{t-k_2}\) with \(I_{t-k_3}\), and so on until \(I_{t-k_n}\) is combined with \(I_t\). This incremental fusion reduces computational overhead during inference, as the intermediate results from closer frames can be reused for subsequent target frames, streamlining the process.
}
	
	\revision{In particular, we first apply a P2T pooling layer \cite{wu2022p2t}, denoted as \(f(\cdot)\), to enhance each intermediate feature map \(U_{t_i}^l\), formulated as:
\begin{equation}
    D_{t_i}^l = f(U_{t_i}^l).
\end{equation}

After obtaining the enhanced feature set \(\{D_{t_i}^l \}_{l=1}^L\), we introduce an attention mechanism to extract important information from frames and propagate it through the temporal sequence.

For simplicity, we omit the superscript \(l\) in \(D_{t_i}^l\). The first frame feature, \(D_{t_0}\), is directly treated as the representative feature for the final decoder input, where \(D_{\text{past}} = D_{t_0}\). For the second frame feature, \(D_{t_1}\), we derive three feature maps using fully-connected layers:

\begin{equation}\label{eq:qkv}
    Q_{t_i} = {\rm FC}(U_{t_i}), \quad
    K_{\text{past}} = {\rm FC}(D_{\text{past}}), \quad
    V_{\text{past}} = {\rm FC}(D_{\text{past}}),
\end{equation}
where \({\rm FC} (\cdot)\) represents the fully-connected layer. We then compute the attention map between two frames, along with the intermediate feature \(N_{t_i}\), as follows:

\begin{equation}{\label{equ:4}}
    \begin{aligned}
        & A_{t_i} = Q_{t_i} \times K_{\text{past}}, \\
        & N_{t_i} = A_{t_i} \times V_{\text{past}}.
    \end{aligned}
\end{equation}

Here, \(A_{t_i} \in \mathbb{R}^{HW \times HW}\) and \(N_{t_i} \in \mathbb{R}^{HW}\). The feature \(D_{\text{past}}\) is updated as \(N_{t_i}\), and for subsequent frames, we follow the same process to generate an attention map. At each scale, every frame produces an attention map using the previous historical feature \(D_{\text{past}}\).

	However, these attention maps are generated without exchanging information across different scales. For instance, shallow layer features offer fine-grained but less semantic information, while deeper layers capture more semantic but coarser details. This highlights the need for a method to enable cross-scale information communication.}

\revision{Inspired by \cite{sun2022mining}, we introduce a multi-scale affinity aggregation module for frame \(t_i\), which aggregates attention maps from deeper layers to shallower ones. This can be formulated as:}

	%对于backbone的输出，我们分别有四个尺度
	%[1,1,225,512],[1,1,900,320],[1,1,3600,128],[1,1,14400,64]
	%拿第一个为例子
	%原始特征[1,1,225,512] p2t之后[1,1,512,330]
	%qk attention得到的affinity [1,1,225,330]
	%qk*v [1,1,225,330] [1,1,330,512] = [1,1,225,512]
	% 
	%为什么能concat？
	%对于四个维度的特征，我们分别有经过p2t之后的特征
	%[1,1,512,330],[1,1,320,330],[1,1,128,330],[1,1,64,330]
	%现在我们就可以在c维上concat了
	\begin{equation}\label{eq:affity}
		\begin{aligned}
			& B_{t_i}^L = A_{t_i}^L, \\
			& B_{t_i}^l = {\rm Conv}({\rm Upsample}(B_{t_i}^{l+1}+A_{t_i}^l), l = L-1,...,1.
		\end{aligned}
	\end{equation}
	\revision{where \(L\) denotes the layer depth, and \(B_{t_i}^l\) represents the refined attention map. Since the attention map \(B_{t_i}^l\) is refined across multiple scales, it encapsulates the informative affinity among multi-scale features. Using this refined attention \(B_{t_i}\) along with the corresponding value \(V_{\text{past}}\), the fused features can be computed as:
}
	\begin{equation}\label{eq:temporal}
		O_{t} = B_{t_i} \times V_{past}.
	\end{equation}
	% By assigning $D_{past} = O_{t_i}$, we can use \ref{eq:qkv} - \ref{eq:temporal} to further aggregate $D_{past}$ and the feature map from the frame $I_{t_{i+1}}$. 
	Finally, we apply an auxiliary convolution head $\phi_{seg}$ to compute a cross-entropy loss:
\begin{equation}\label{eqa:auxiliary}
    \mathcal{L}_{aux} = {\rm CE}(\phi_{seg}(O_t),G_t),
\end{equation}
in which ${\rm CE}(\cdot)$ is the standard cross-entropy loss function.

	\subsection{Random Frame Enhancement}\label{sec:random frame}
        To capture long-term dependencies among video frames, a straightforward approach would be to input all frames into the model. However, this would be computationally prohibitive. Instead, we adopt a strategy of randomly selecting a temporally distant frame relative to the target frame. By aggregating information from this randomly selected frame into the target frame's features, we enable the model to develop a more comprehensive understanding of the environment, thereby enhancing model training.

       \revision{ It is worth noting that our STCF module from \cref{sec:stcf} integrates features from two closely positioned frames during each operation, gradually incorporating information from neighboring frames. In this context, the random frame selection complements the STCF module. Since the environment depicted in the randomly selected frame may differ from that of the target frame, we employ an attention mechanism to automatically emphasize relevant information while reducing noise.
}

    \begin{figure}[!t]
        \centering
        \setlength{\abovecaptionskip}{0.6cm}
        \includegraphics[width=.6\linewidth]{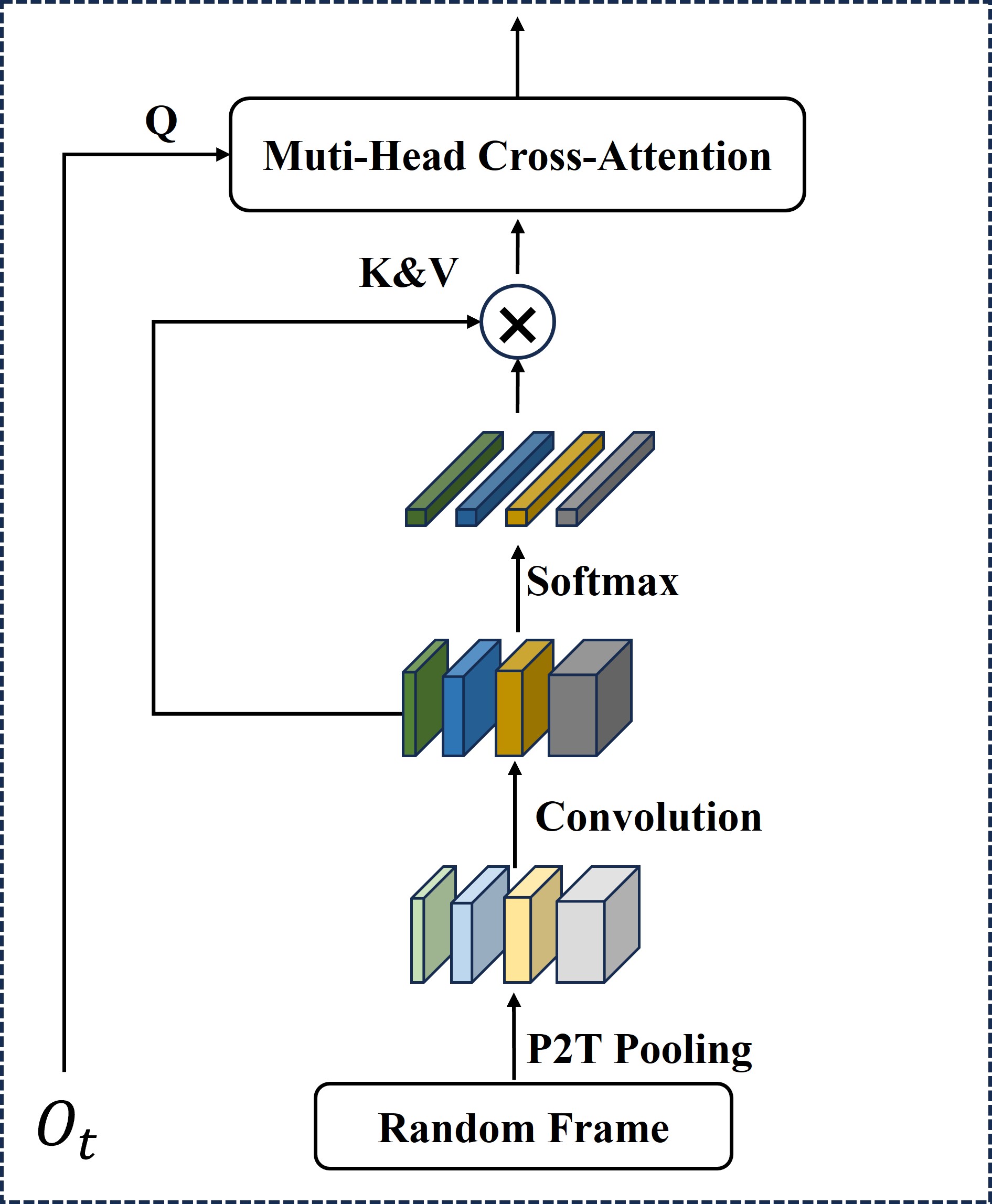}
        \vspace{-.10in}
        \caption{The architecture of the Random Frame Enhancement module utilizes cross-attention to aggregate contextual information from the randomly selected frame.}
        \label{fig:random-frame}
        \vspace{-.10in}
    \end{figure}

	Suppose that the backbone features of the random frame $I_{random}$ from a P2T \cite{wu2022p2t} pooling layer are denoted as $D_{random}^l$ for the scale $l$. We aggregate these multi-scale features for the scales from $l$ to $L$ using
	%改2
	\begin{equation}\label{eq:8}
		\begin{aligned}
			& D_{random} = {\rm Concat} [D_{random}^l,D_{random}^{l+1}...D_{random}^L], \\
			& D_{random} = {\rm Linear}(D_{random}). \\
		\end{aligned}
	\end{equation}

\revision{Inspired by the Object Contextual Representation (OCR) module \cite{yuan2020object}, we propose a contextual representation method that characterizes each pixel by utilizing its corresponding contextual information.

We begin by applying a simple convolution to extract the region-level representation. Next, we sum the pixel representations, weighted by their association with the object region. Using an aggregation module, we compute the relationships between pixels and object regions to derive enhanced context-level information, as follows:
}

	\begin{equation}\label{eq:9}
		\begin{aligned}
			& L_{d} = {\rm Conv}(D_{random}), \\
			& L_r = {\rm Softmax}(L_d) \times D_{random}.
			% & L_{p_i} = \frac{r^jp_i}{\sum_{j=1}^Kr^{j}} \\ 
		\end{aligned}
	\end{equation}
	Then, for the target frame, we have the short-term temporal representation $O_t$ and the long-term context representation $L_{r}$. Next, we apply a multi-head self-attention mechanism to aggregate these features:
	%说明根号c
    \begin{equation}\label{eq:10}
        \begin{aligned}
            & Q = {\rm FC}(O_{t}), K = {\rm FC}(L_r), V = {\rm FC}(L_r), \\
            & \hat{O}_t = {\rm Softmax}\left(\frac{Q \times K^T}{\sqrt{C}}\right) \times V, \\
        \end{aligned}
    \end{equation}
    Now, we have the random frame enhanced feature $\hat{O}_t$, where $C$ denotes the feature dimension.

	\subsection{Video Text Encoding Module}
\revision{Given an image \(I \in \mathbb{R}^{ H \times W \times 3}\) and a set of candidate class category prompts \(\{T(n)\}\) for \(n = 1, \dots, N\), instead of directly using the pretrained text encoder from the CLIP model \cite{radford2021learning}, we enhance its performance for our task by introducing a video-specific prompt scheme.  

The core idea is to leverage video context to enrich text representations within the video. A common approach is to directly use the text representation. However, we demonstrate that this can reduce segmentation performance in video contexts. To address this, we introduce a video-specific text encoding module.

In human video comprehension, key terms often help distinguish visual content. For example, including the prompt "water" makes it easier to differentiate between "fish" and "horse" within the visual context.
}
	
	Specifically, we propose the utilization of a multi-head self-attention (MHSA) mechanism \cite{vaswani2017attention} to model the correlation between video clips and text. 
	% To achieve better-aligned features, we follow the approach in \cite{zhou2022extract} and modify the last layer of the image encoder. 
	Given the CLIP image encoder $\Phi^V$ and the text encoder $\Phi^T$, we initially extract dense image features $F^V = \Phi^V(I) \in \mathbb{R}^{ H \times W \times C}$ for each input frame $I \in [I_{t-k_1}, I_t]$ and text features $F^T = \Phi^T(T_n) \in \mathbb{R}^{N \times C}$ for $n \in \{1, \cdots, N\}$.
	We refine the text feature $F^T$ with the visual feature $F^V$ using
	\begin{equation}\label{eq:11}
		\begin{aligned}
			\overline{F^T} = F^T + {\rm MHSA}(F^T,F^V).
		\end{aligned}
	\end{equation}
	%btpc [1,111,80,512] bchw [1,512,24,24]
	After obtaining refined text feature $\overline{F^T} \in \mathbb{R}^{N \times C}$ and image feature $F^V \in \mathbb{R}^{H \times W \times C}$, inspired by \cite{cho2023cat}, we compute a cost volume $ X \in \mathbb{R}^{N \times H \times W}$ by cosine similarity. Formally, it can be defined as:
	\begin{equation}\label{eq:12}
		X=\frac{\overline{F^T} \cdot F^V}{||\overline{F^T}|| \cdot ||F^V||}.
	\end{equation}
	\revision{After constructing the cost volume, we apply a simple convolution on \(X\) to refine it, resulting in the enhanced cost volume \(\tilde{X}\).
}
	
	Since CLIP\cite{radford2021learning} is pretrained on a single image but our model is working with a pixel-level task, it is vital to lead the CLIP to learn the position information. In order to provide corresponding positional information for cost volume, we used shallow features $U_t^{l=1}$ from the backbone, as it contains a large amount of object positional information. We integrated cost volume location data by concatenate it with the existing location information, like
	%说清楚操作办法
	%【B*T，C，H，W】【B*t,c,h,w]
	% 111,128,24,24 111,80,15,15
	% 111，208，24，24
	\begin{equation}\label{eq:13}
		\begin{aligned}
			% & U_t^{l=1} = {\rm repeat}(U_t^{l=1}) \\
			\hat{X} = {\rm Linear}(({\rm Concat}[\tilde{X}, {\rm Upsample}(U_{t}^{l=1}))]. \\
		\end{aligned}
	\end{equation}
	After obtaining the position-acquired cost volume $\hat{X}$, we first perform a upsample operation on cost volume $\hat{X}$. We then concatenate the cost volume $\hat{X}$ and refined features $\hat{O}_t$ from \cref{sec:random frame} on the channel dimension for final prediction. This process can be expressed as 
	%111，128,24,24   111，30，72，72
	\begin{equation}
		\begin{aligned}
			% & \check{X} = {\rm Upsample}(\hat{X})(:,:{\rm Dim}(\hat{O}_t)), \\
			& \check{X} = {\rm Concat}[{\rm Upsample}(\hat{X}), \hat{O}_{t}].
		\end{aligned}
	\end{equation}
	% where ${\rm Dim}(\hat{O}_t)$ denotes the channel dimension of $\hat{O}_t$, ${\rm Upsample}(\hat{X})(:,:{\rm Dim}(\hat{O}_t)) \in \mathbb{R}^{HW \times d}$, $d$ is the channel dimension of $\hat{O}_t$.
    After that, we use a convolution head to obtain the prediction result. Finally, the training loss is calculated by a standard CE loss and an auxiliary loss from \cref{eqa:auxiliary}, like:
	\begin{equation}\label{eq:loss}
		\begin{aligned}
			& \mathcal{L}_{main} =  {\rm CE}(\phi_{seg}(\Check{X}),G_t), \\ 
			& \mathcal{L} = \alpha\mathcal{L}_{main} + \beta\mathcal{L}_{aux}.
		\end{aligned}
	\end{equation}
	where $\alpha$ and $\beta$ are trade-off parameters.

\begin{table*}[!t]
    \centering
    \begin{minipage}[t]{0.48\linewidth}
        \centering
        \caption{The results on the  VSPW \cite{miao2021vspw}dataset. }
        \setlength{\tabcolsep}{4pt} % 设置列间距
        \begin{tabular}{llllll}
            \hline\noalign{\smallskip}
            \rowcolor{headerColor}
            Methods & Backbone & mIoU$\uparrow$	& fwIoU$\uparrow$ & mAcc$\uparrow$ & pAcc$\uparrow$ \\
            \noalign{\smallskip}
            \hline
            \noalign{\smallskip}
            Zegformer \cite{ding2022decoupling} & ResNet-101 & 1.91 & 4.91 & 6.93 & 12.21 \\
            DeOP \cite{Han2023ZeroShotSS} & ResNet-101 & 2.58 & 6.29 & 9.33 & 17.68 \\
            SAN \cite{xu2023side} & ResNet-101 & 6.56 & 12.9 & 16.45 & 24.33 \\
            zsseg.baseline \cite{xu2021} & ResNet-101 & 9.46 & 16.89 & 20.97 & 30.21 \\
            FreeSeg \cite{qin2023freeseg} & ResNet-101 & 13.13 & 22.73 & 29.66 & 33.11 \\
            Ours & ResNet-101 & \textbf{17.22} & \textbf{31.89} & \textbf{53.22} & \textbf{50.91} \\
            \hline
            SAM \cite{kirillov2023segment} & ViT-B2 \cite{dosovitskiy2020vit} & 12.67 & 22.43 & \textbf{47.32} & 33.11 \\
            Ours & ViT-B2 & \textbf{17.99} & \textbf{25.22} & 34.07 & \textbf{41.92} \\
            \hline
        \end{tabular}
        \label{table:compare}
    \end{minipage}%
    \hfill
    \begin{minipage}[t]{0.48\linewidth}
        \centering
        \caption{The results on the Cityscapes\cite{cordts2016cityscapes} dataset. The results are obtained by directly testing on the dataset using the model pretrained on  VSPW \cite{miao2021vspw} dataset.}
        \setlength{\tabcolsep}{4pt} % 设置列间距
        \begin{tabular}{llllll}
            \hline\noalign{\smallskip}
            \rowcolor{headerColor}
            Methods & Backbone & mIoU$\uparrow$ & fwIoU$\uparrow$ & mAcc$\uparrow$ & pAcc$\uparrow$ \\
            \noalign{\smallskip}
            \hline
            \noalign{\smallskip}
            DeOP \cite{Han2023ZeroShotSS} & ResNet-101 & 3.47 & 5.94 & 10.04 & 22.22 \\
            FreeSeg \cite{qin2023freeseg} & ResNet-101 & 8.37 & 3.19 & 16.96 & 6.07 \\
            SAN \cite{xu2023side} & ResNet-101 & 14.68 & 28.5 & 27.59 & 37.28 \\
            Ours & ResNet-101 & \textbf{27.65} & \textbf{41.12} & \textbf{71.29} & \textbf{64.88} \\
            \hline
            SAM \cite{kirillov2023segment} & ViT-B2 \cite{dosovitskiy2020vit} & 11.09 & 20.02 & 43.13 & 29.77 \\
            Ours & ViT-B2 &  \textbf{26.56} & \textbf{39.31} & \textbf{63.77} & \textbf{57.91}\\
            \hline
        \end{tabular}
        \label{tab:cityscapes}
    \end{minipage}
\end{table*}

    \subsection{Complexity Analysis}
 \revision{Here we analyze the complexity of our methods for processing video clips $([I_{t-k_1}, \cdots, I_{t-k_n}, I_{random}])$. The Spatial-Temporal Context Fusion Module has a complexity of \(\mathcal{O}((n-1)(hwc^2+h^2w^2c))\), where \(h\), \(w\), and \(c\) represent the height, width, and number of channels in the input frames, respectively. The complexity of the Random Frame Enhancement module has a complexity of \(\mathcal{O}(hwc^2+h^2w^2c)\). For the Video Text Encoding Module, it has a complexity of \(\mathcal{O}(n(hwc^2+h^2w^2c))\).}
	
	\section{Experiments}
	
	\subsection{Datasets}
	We conduct out experiments on two popular datasets: Video Scene Parsing in the Wild (VSPW) \cite{miao2021vspw} and CityScapes \cite{cordts2016cityscapes}. 
	
	\para{VSPW.} \revision{VSPW \cite{miao2021vspw} provides a large-scale benchmark with dense annotations for well-trimmed, long-term video clips, making it one of the most challenging datasets for video semantic segmentation tasks. Its diversity of 124 classes makes it ideal for evaluating open-vocabulary video segmentation. The dataset contains 2806/343/387 videos in the training, validation, and test sets, respectively, with 198244/24502/28887 frames. Each video has an average of 71 frames, with a maximum of 482, resized to 480$\times$853. Our OV-VSS model will be trained on a subset of classes and tested on another.
}
	
	\para{CityScapes.} \revision{CityScapes \cite{cordts2016cityscapes} is a widely recognized dataset in the field of video semantic segmentation and has been a key benchmark in earlier research. Notably, for each video, only the 20th frame is annotated at the pixel level. The dataset consists of 5,000 labeled frames, divided into 2975 for training, 500 for validation, and 1525 for testing. This setup makes CityScapes particularly suitable for evaluating models with sparse annotations across video sequences.
}
	\subsection{Experimental Setting}
	Similar to existing work, we adopt ResNet \cite{he2016deep} as the backbone. We set the temporal distances between adjacent frames and the target frame to 3, 6, and 9.
	We initialize the backbone ResNet with ImageNet pretrained weights, and other parts of the model randomly. We train OV-VSS on VSPW for 10,0000 iterations with a batch size of 4, learning rate initialized as 3e-6. We employ AdamW \cite{kingma2014adam} with weight decay as 1e-2. We first train the model with a linear warm-up for 1500 iterations. We use 1 NVIDIA RTX 3090 GPU for training, costing 20 hours.
	
	For the pretrained CLIP text encoder, the ViT-B/16 \cite{dosovitskiy2020vit} backbone is used as default if not specified. We also use prompt templates with class names for all experiments. We follow \cite{ding2022decoupling} to generate multiple text embeddings for each class name based on several text prompt templates. Then, we use the mean of these embeddings for final classification.
	
	During training, we adopt random scaling and random cropping for data augmentation. In testing, we perform a single-scale test and use the original 480p image size for inference.

	\subsection{Evaluation Protocol.} Following the literature of video semantic segmentation \cite{sun2022mining,sun2022coarse}, we mainly conduct experiments on VSPW datasets \cite{miao2021vspw}, which containing 124 classes. We evaluate our model on their validation set containing 24392 images. 
	
	We first compare the State-of-the-Art results on the VSPW dataset. As there are no existing video-based open vocabulary semantic segmentation methods for comparison, we conduct a comparison between image-based open vocabulary methods and our approach. By splitting video clips into continuous frames, we train all models and evaluate them on the test set.
	
	In order to verify the open-vocabulary capability of the model, previous work \cite{ding2022decoupling} requires careful selection of visible classes during training and invisible classes during testing according to super categories. We believe that such a setting limits the ability to fully validate the model. To demonstrate the superiority of our model's performance, we directly select the first 80 classes as visible classes during training and the remaining 44 classes as invisible classes during testing, following the category order of VSPW. To prevent the model from learning information about invisible classes during training, we mask out the invisible classes based on their classification, ensuring that the model only learns the features of the predefined visible classes. Furthermore, we mask out the invisible classes in the video frames before feeding them into the model to ensure that the model solely focuses on the visible classes we set.
	
	To verify our cross-dataset performance, we also conducted additional experiments. We directly validated the model trained on the top 80 visible classes in VSPW \cite{miao2021vspw} on the CityScapes \cite{cordts2016cityscapes} dataset. After expert confirmation, 8 categories in CityScapes were new categories that the model had not encountered before.
	
	For evaluation, we adopt four metrics to evaluate the results.
	\begin{enumerate}
		\item mIoU: Mean Intersection over Union (mIoU). mIoU is a standard metric for evaluating image segmentation models. It calculates the average overlap between predicted and ground truth segmentation regions across all classes, balancing precision and recall.
		\item FWIoU: Frequency Weighted Intersection over Union. FWIoU evaluates segmentation performance by weighting the IoU of each category by its pixel count, emphasizing classes with more pixels and accounting for class distribution.
		\item mAcc: Mean Accuracy. mAcc reports the average accuracy for each class in a dataset, ensuring equal contribution from each class. It is useful for datasets with imbalanced class distributions.
		\item pAcc: Pixel-wise Classification Accuracy. pAcc measures the proportion of correctly classified pixels in an image. It provides an overall accuracy assessment but can be biased toward larger classes.
	\end{enumerate} 

	\subsection{Open-Vocabulary Semantic Segmentation Results}
	\para{Image-based methods.} For the image-based methods, Zegformer \cite{ding2022decoupling} decouples segmentation and classification by generating class-agnostic segment masks. DeOP \cite{Han2023ZeroShotSS} proposes a decoupled one-pass structure for computational efficiency. SAN \cite{xu2023side} frames the segmentation task as a region recognition problem by attaching a side network. zsseg.baseline \cite{xu2021} presents a two-stage approach to address the discrepancy in processing granularity. FreeSeg \cite{qin2023freeseg} introduces a generic framework for universal open-vocabulary image segmentation.

    \para{Segment-Anything-Model.} Given the lack of text prompt functionality in SAM \cite{kirillov2023segment}, we designed a novel approach that combines CLIP \cite{radford2021learning} with the SAM model. This approach aims to establish a baseline for SAM’s performance by leveraging CLIP's ability to understand both text and images. Specifically, our method utilizes SAM models to identify regions of interest in video images, after which CLIP is used to classify each identified region into a predefined class. This allows us to adapt SAM for Open-Vocabulary Video Segmentation, enabling the segmentation of novel classes without additional training.
	
	% Since SAM \cite{kirillov2023segment} did not release the capability of text prompt, we have designed a method that combines CLIP \cite{radford2021learning} with the SAM model, which can verify the baseline of the SAM model. The specific method is to first use the SAM model to identify the region of interest on the image, and then use the CLIP to distinguish whether the region corresponds to the content of the text.

	% In the realm of image-based methods, Zegformer \cite{ding2022decoupling} introduces a novel approach by decoupling segmentation and classification processes, thereby generating class-agnostic segment masks to enhance overall performance. Similarly, DeOP \cite{Han2023ZeroShotSS} proposes a decoupled one-pass structure aimed at optimizing computation efficiency within the segmentation framework. Another noteworthy method, SAN \cite{xu2023side}, adopts a distinctive approach by treating the segmentation task as a region recognition problem, achieved through the integration of a side network. Meanwhile, zsseg.baseline \cite{xu2021} addresses granularity discrepancies in processing through a two-stage methodology, enhancing the overall segmentation process. Additionally, FreeSeg \cite{qin2023freeseg} presents a versatile framework designed to facilitate universal open-vocabulary Image Segmentation, marking a significant advancement in the field.

	\subsection{Segmentation Results}\label{sub4.1}
	In this experimental analysis, we begin by training the model using the initial 80 classes for training purposes. During testing, we evaluate its performance on the remaining 44 novel classes. We categorize the methods into two groups based on their backbone architectures.

    Regarding segment IoU, our approach outperforms other methods, surpassing the second-best methods by 4.09\% and 5.32\%, respectively. The state-of-the-art comparisons on the VSPW dataset \cite{miao2021vspw} are presented in \cref{table:compare}.
    
    To validate the zero-shot capability of our model on novel categories, we conduct direct testing on the CityScapes dataset \cite{cordts2016cityscapes} without any modifications. The results, showcased in \cref{tab:cityscapes}, underscore the superior performance of our method across both ResNet-101 and ViT backbones.
    
\begin{figure*}[!t]
		\centering
		\includegraphics[width=\linewidth]{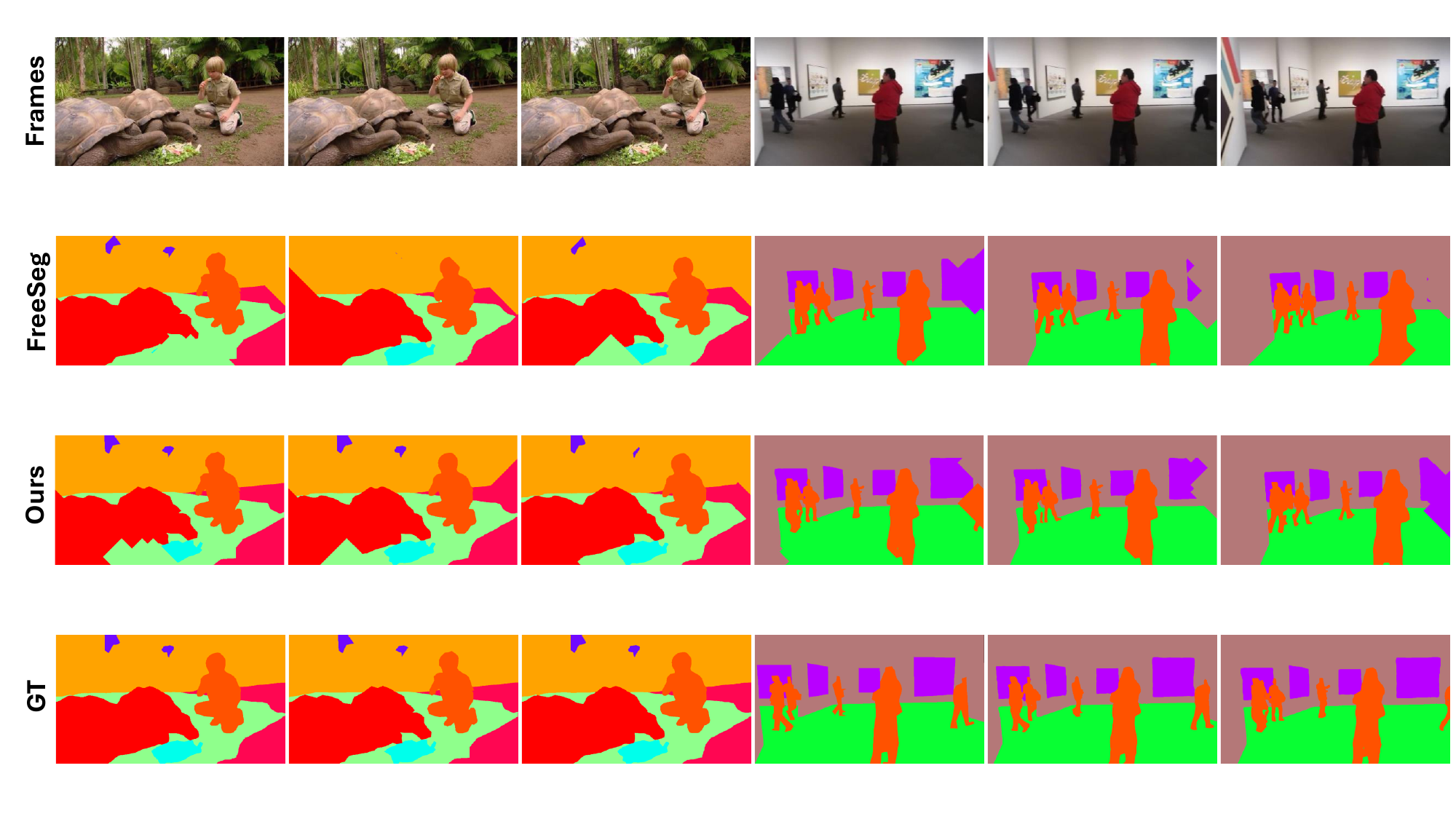}
		\caption{Qualitative results on the VSPW dataset. From top to bottom: Original frames, segmentation results from the image-based method (FreeSeg), segmentation results from our method (ViT-B), and the ground truth mask. Clearly, our model generates better segmentation results.}
		\label{fig:seg_result}
	\end{figure*}

    In addition to the quantitative comparisons mentioned above, we also perform qualitative assessments by juxtaposing our proposed method with the baseline on sampled video clips, as illustrated in \cref{fig:seg_result}. Upon closer inspection, our method consistently generates more accurate segmentation masks across the sampled video clips, reaffirming its efficacy and robustness in real-world scenarios.
    
        \begin{figure*}[!t]
		\centering
		\includegraphics[width=.95\linewidth]{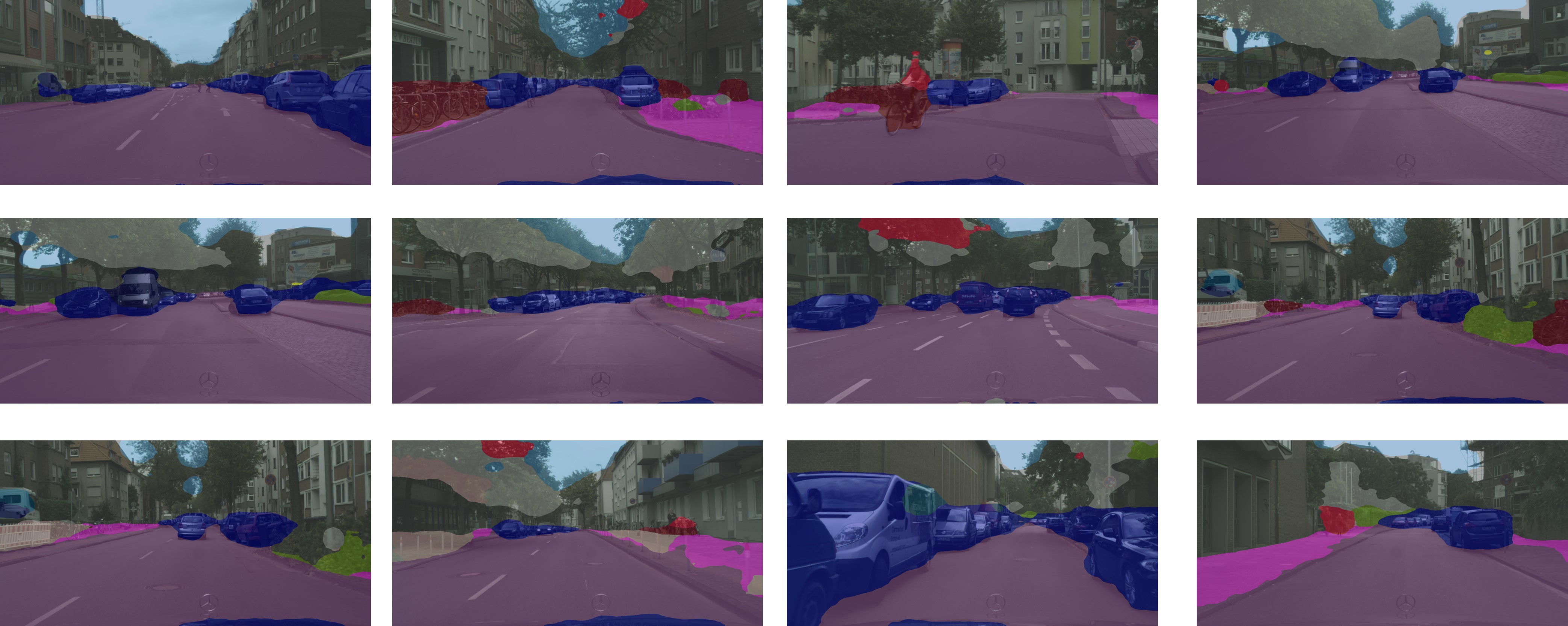}
		\caption{Qualitative results on CityScapes datasets. We conduct inference on Cityscapes using the model trained on VSPW without further training. Although we have achieved good results on this dataset, our model still cannot accurately segment complex image regions, such as when trees and sky overlap.}
		\label{fig:seg_cityscapes_result}
	\end{figure*}

	\subsection{Ablation Study}
	
	We conduct ablation studies on the large-scale VSPW \cite{miao2021vspw} dataset to validate the key designs of OV-VSS and report the results in~\cref{table:combined-ablation}. For fairness, we maintain the same settings as in \cref{sub4.1}, unless otherwise specified. The ablation studies are conducted on the ViT-B2 backbone.

    \textit{1) Ablation Study on Key Components:} Results shows the ablation study for each component of OV-VSS. First, we establish a simple baseline that aligns image features with text features. Initially, we train the model in a fully-supervised manner, which is GPU-intensive and requires per-image labels. By integrating a random frame enhancement module, we achieve better results (+0.91\%) with a resource-friendly training mechanism. This method processes frames one by one but loses the temporal information between frames. By utilizing a spatial-temporal context fusion module, we observe an accuracy improvement of 0.43\%. Lastly, since CLIP is trained on image-text pairs, which may degrade performance on video-text tasks, we propose a video-based text encoding mechanism that yields an impressive 1.67\% performance boost.
    
    \textit{2) Ablation Study on Fusing Strategy:} We also evaluate the performance of our method with respect to different temporal fusion strategies. We make the following observations: First, using a simple element-wise addition results in 5.42\% mIoU, possibly because simple addition neglects crucial information and diminishes the informative features. Next, by employing a concatenation-based fusion strategy, we achieve 17.99\% mIoU.
	
	\textit{3) Ablation Study on Random Frame Enhancement:}
	% On \ref{table:global_category_context}, we analyze the random frame enhancement module in the following ways. First, we removed the introduction of random frame, we obtained the result of 13.55\% mIoU. Then, we try to train the model under a a-label-per-image condition, and compute loss on each frame, we achieve 12.55 \% mIoU. 
	We also study the random frame enhancement module through several methodologies. Initially, we conducted an experiment wherein the introduction of the random frame was deliberately omitted. This adjustment yielded a segmentation result of 13.55\% mIoU. Following this, we proceeded to train the model under a condition referred to as ``a-label-per-image'', where the model was trained with a single label per image. Under this condition, loss computation was performed individually on each frame, resulting in a segmentation performance of 12.55\% mIoU.
	
	\textit{4) Ablation Study on the Video Text Encoding Module:}
	% On \ref{table:text_encoding}, we studied the effectiveness of our video text encoding module below. First, we only retained a multi-head self-attention structure to model the relationship between text and images, and obtained the mIoU of 14.88\% mIoU. Then, our current form uses the FFN structure and obtained the mIoU of 17.99\% mIoU.
	We  further delve into the efficacy of our video text encoding module through a series of evaluations. Initially, we conducted an experiment wherein we retained solely a multi-head self-attention structure to model the intricate relationship between text and images. This initial configuration yielded a segmentation performance of 14.88\% mIoU. Subsequently, we integrated the FFN structure into the encoding module, aiming to enhance the model's capability in capturing text-image interactions. With this enhancement, we observed a notable improvement in segmentation performance, achieving an mIoU of 17.99\%.
	
	\textit{5) Ablation Study on the loss ratio:}
	% On \ref{table:text_encoding}, we studied the effectiveness of our video text encoding module below. First, we only retained a multi-head self-attention structure to model the relationship between text and images, and obtained the mIoU of 14.88\% mIoU. Then, our current form uses the FFN structure and obtained the mIoU of 17.99\% mIoU.
	\revision{As detailed in~\cref{table:combined-ablation}, we assessed the effectiveness of the loss ratio through a series of experiments. Initially, we focused on adjusting the ratio between the primary and auxiliary losses. In the first experiment, we set \(\alpha\) to 0.1 and \(\beta\) to 1, achieving an mIoU of 14.88\% on the test set. Increasing \(\alpha\) to 1 resulted in a notable performance boost, with an mIoU of 17.99\%. However, further increasing \(\alpha\) to 10 led to a decline in accuracy, reducing the mIoU to 15.43\%. In addition to this, we have further investigated the impact of modifying $\beta$ alone on the segmentation accuracy, while keeping $\alpha$ at 1 and reducing $\beta$ to 0.1, the segmentation result was 15.38\%. Keeping $\alpha$ at 1 and reducing $\beta$ to 0.1, the segmentation result was 13.78\%.}

\section{Future Work}
\revisionminor{
While this work presents a strong foundation for Open Vocabulary Video Semantic Segmentation (OV-VSS), there are several avenues for future research and improvements that could further advance the field:

\begin{itemize}
\item \textbf{Exploring other Multi-Modal Learning:} 
Future work could explore integrating additional modalities, such as infrared images~\cite{zhang2021deep}, depth images~\cite{song2017semantic}, or point clouds~\cite{tang2022contrastive}, to enhance segmentation performance. Leveraging these diverse data sources could improve the model's robustness and accuracy, especially in challenging environments where visual information alone is insufficient.

\item \textbf{Addressing Label Noise in Training and Testing:} 
In real-world applications, not only might the categories of test samples be unclear, but training labels can also be inaccurate due to errors or inconsistencies in the annotation process~\cite{song2022learning,zhang2024cognition}. Future work should investigate how our method handles such label noise, particularly when training labels are noisy. While our current approach demonstrates robust performance with clean labels, it is important to explore how the model performs when the training data contains noisy labels, as this is a common scenario in practical deployments.

\item \textbf{Improving Performance with Low-Quality Input Data:} 
In scenarios where the input data quality is low, applying image enhancement techniques could improve the model's performance. Future work should explore how pre-processing with enhancement methods~\cite{li2023feature,li2022all} affects segmentation accuracy and investigate joint optimization approaches to improve both data quality and segmentation results simultaneously.

\item \textbf{Few-Shot Segmentation Capabilities:} 
   While our method performs well in the open vocabulary setting, the ability to handle few-shot learning~\cite{wang2019panet,zhang2024part} remains an important avenue for future work. Investigating whether the model can further improve performance in scenarios with limited annotated data is crucial. 

   \end{itemize}

By addressing these challenges, we believe that OV-VSS could evolve into a more powerful and versatile tool for video analysis, with a broad range of real-world applications, from autonomous systems to interactive media editing.}

\begin{table}[!t]
    \centering
    \caption{ Various ablation studies on the large-scale VSPW \cite{miao2021vspw} dataset. }
    \setlength{\tabcolsep}{6pt} % 调整列间距
    \begin{tabular}{p{3.5cm}|p{3cm}|c}
        \hline\noalign{\smallskip}
        \rowcolor{headerColor}
        \multicolumn{2}{c|}{Ablation Setting} & mIoU(\%) \\
        \noalign{\smallskip}\hline\noalign{\smallskip}
        \multirow{4}{*}{Key Component}
            & Baseline  w/o text  & 14.98 \\
            & + Random Frame Enhancement & 15.89  \\
            & + Spatial-Temporal Context Fusion & 16.32 \\
            & + Video Text Encoding & \textbf{17.99}  \\
        \hline
        \multirow{2}{*}{Fusing Strategy}
            & Element-Wise Addition & 5.42 \\
            & Concat on Channel & \textbf{17.99} \\
        \hline
        \multirow{2}{*}{Random Frame Enhancement}
            & 4 frames, 4 annotations & 12.25 \\
            & 4 frames, 1 annotation & 13.55 \\
            & 4 frames + 1 random frame, 1 annotation & \textbf{17.99} \\
        \hline
        \multirow{2}{*}{Video Text Encoding}
            & Self-Attention & 14.88 \\
            & Self-attention \& FFN & \textbf{17.99} \\
        \hline
        \multirow{3}{*}{Loss Hyperparameters}
            & $\alpha=0.1$, $\beta=1$ & 14.88 \\
            & $\alpha=1$, $\beta=1$ & \textbf{17.99} \\
            & $\alpha=10$, $\beta=1$ & 15.43 \\
            & $\alpha=1$, $\beta=0.1$ & 15.38 \\
            & $\alpha=1$, $\beta=10$ & 13.78 \\
        \hline
    \end{tabular}
    \label{table:combined-ablation}
\end{table}

    % \subsection{Discussion}
    % \textcolor{red}{The modules we proposed in modules a and b in Chapter 3 have two main advantages. Firstly, our proposed cross frame fusion strategy can effectively reduce computational complexity during inference, as frames with historical time information obtained in the previous sequence can be directly used by subsequent frames. Secondly, our proposed random frame reinforcement module can effectively reduce the annotation of training data, which has also been verified in our validation experiments.}
	
	\section{Conclusion}
	\revision{In this study, we explore the task of Open-Vocabulary Video Semantic Segmentation (OV-VSS) and propose a comprehensive end-to-end framework. Our approach introduces several novel components to address the complexities of OV-VSS. First, we develop a Spatial-Temporal Context Fusion module to effectively capture intra-frame correlations. Building on this, a Random Frame Enhancement module refines the target frame based on these correlations, improving segmentation accuracy. Additionally, we incorporate a video-specific text encoding module. Our framework not only addresses open-vocabulary challenges but also establishes a strong baseline for OV-VSS tasks.}
	
\section*{Acknowledgment}
This work was supported by the Key Program for International Cooperation of Ministry of Science and Technology of China (No.2024YFE0100700) and the National Natural Science Foundation of China (NSFC) under Grant 62020106011.
\bibliographystyle{IEEEtran}
\bibliography{main}

\begin{IEEEbiography}
[{\includegraphics[width=1in,height=1.25in,clip,keepaspectratio]{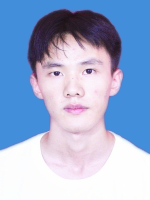}}]
{Xinhao Li} received his B.E. from SouthWest Jiaotong University (SWJTU), Chengdu, China, in 2022. He has been working towards the M.S. degree in Information and Communication Engineering, the University of Electronic Science and Technology of China (UESTC). His research interests include computer vision and cross-modality learning.
\end{IEEEbiography}

\begin{IEEEbiography}
[{\includegraphics[width=1in,height=1.25in,clip,keepaspectratio]{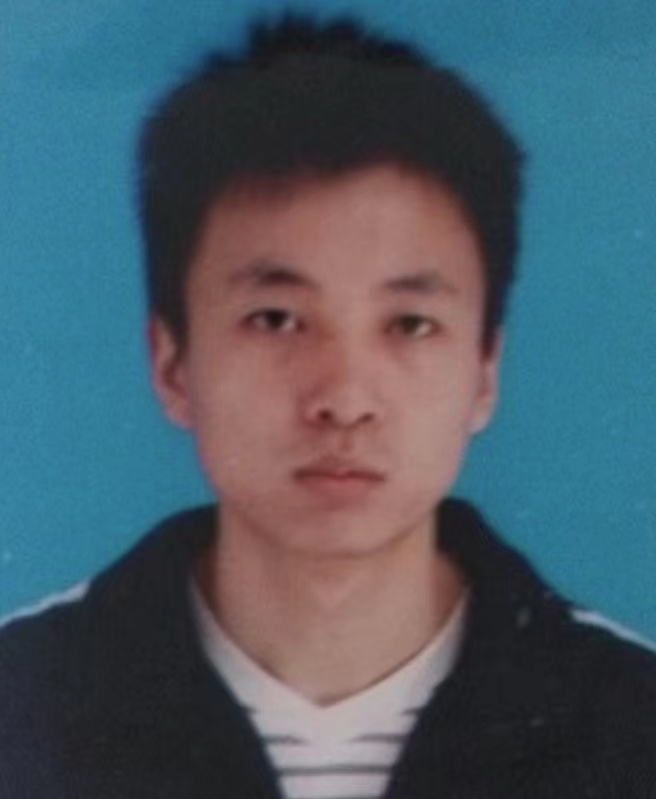}}]
{Yun Liu} received his B.E. and Ph.D. degrees from Nankai University in 2016 and 2020, respectively. Then, he worked with Prof. Luc Van Gool as a postdoctoral scholar at the Computer Vision Lab, ETH Zurich, Switzerland. After that, he worked as a senior scientist at the Institute for Infocomm Research (I2R), A*STAR, Singapore. Now, he is a professor at the College of Computer Science, Nankai University. His research interest is computer vision.
\end{IEEEbiography}

\begin{IEEEbiography}
[{\includegraphics[width=1in,height=1.25in,clip,keepaspectratio]{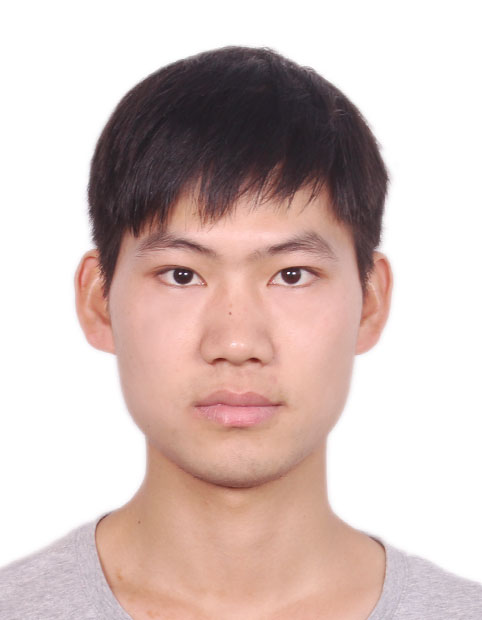}}]
{Guolei Sun} received master’s degree in computer
science from the King Abdullah University of Science
and Technology (KAUST), in 2018, and the PhD
degree from ETH Zurich, Switzerland, in Prof. Luc
Van Gool’s Computer Vision Lab in 2024. Before that, he obtained bachelor's degree from Huazhong University of Science and Technology (HUST). From 2018 to 2019, he worked as a research engineer with the Inception
Institute of Artificial Intelligence, UAE. His research
interests include deep learning for video understanding, semantic/instance segmentation, object counting,
and weakly supervised learning.
\end{IEEEbiography}

\begin{IEEEbiography}
[{\includegraphics[width=1in,height=1.25in,clip,keepaspectratio]{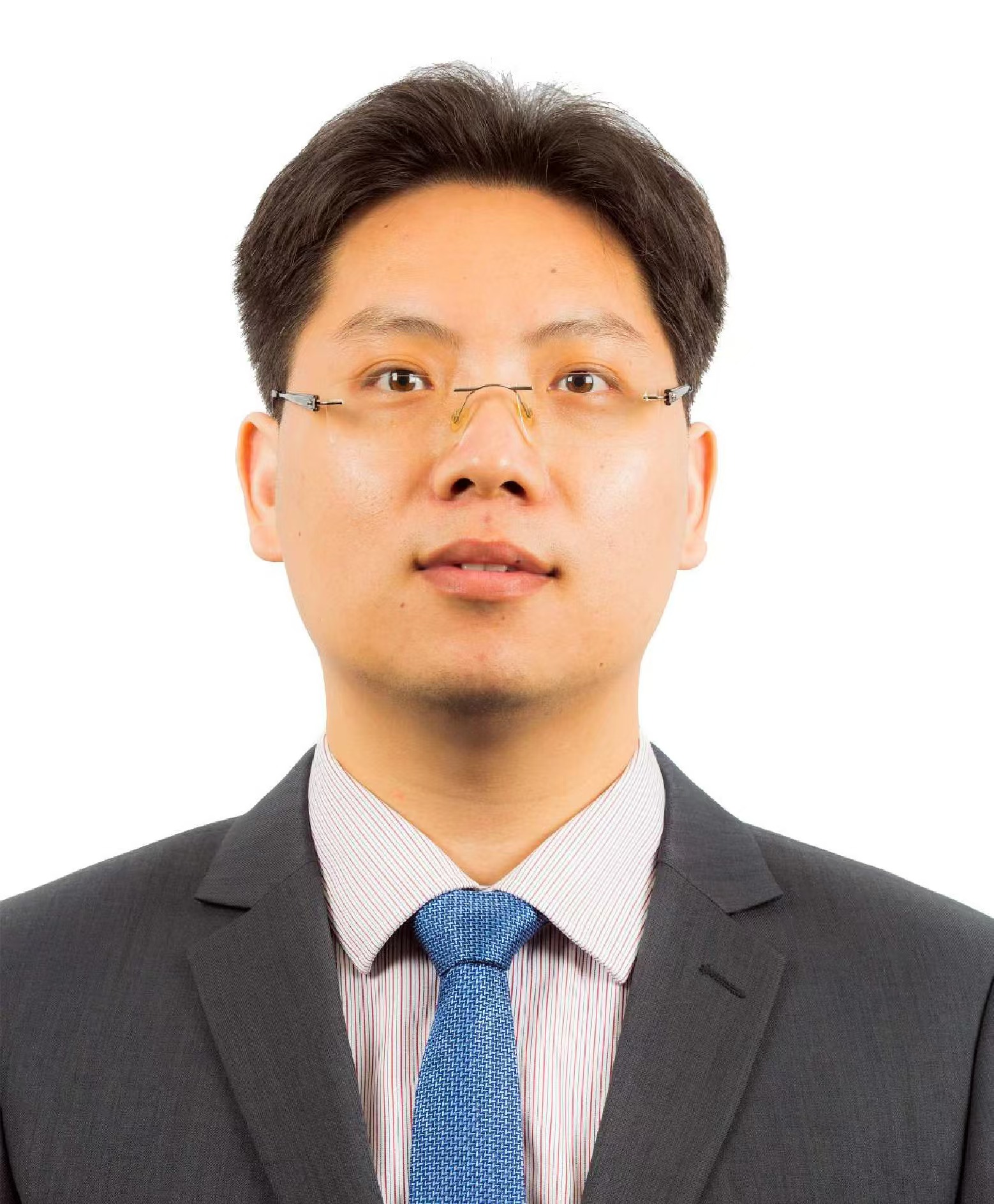}}] {Min Wu} is currently a Principal Scientist at Institute for Infocomm Research (I2R), Agency for Science, Technology and Research (A*STAR), Singapore. He received his Ph.D. degree in Computer Science from Nanyang Technological University (NTU), Singapore, in 2011 and B.E. degree in Computer Science from University of Science and Technology of China (USTC) in 2006. He received the best paper awards in IEEE ICIEA 2022, IEEE SmartCity 2022, InCoB 2016 and DASFAA 2015. He also won the CVPR UG2+ challenge in 2021 and the IJCAI competition on repeated buyers prediction in 2015. He has been serving as an Associate Editor for journals like Neurocomputing, Neural Networks and IEEE Transactions on Cognitive and Developmental Systems, as well as conference area chairs of leading AI and machine learning conferences, such as ICLR, NeurIPS, etc. His current research interests focus on AI and machine learning for time series data, such as deep learning, self-supervised learning, domain adaptation, and knowledge distillation for time series data.
\end{IEEEbiography}

\begin{IEEEbiography}[{\includegraphics[width=1in,height=1.25in,clip,keepaspectratio]{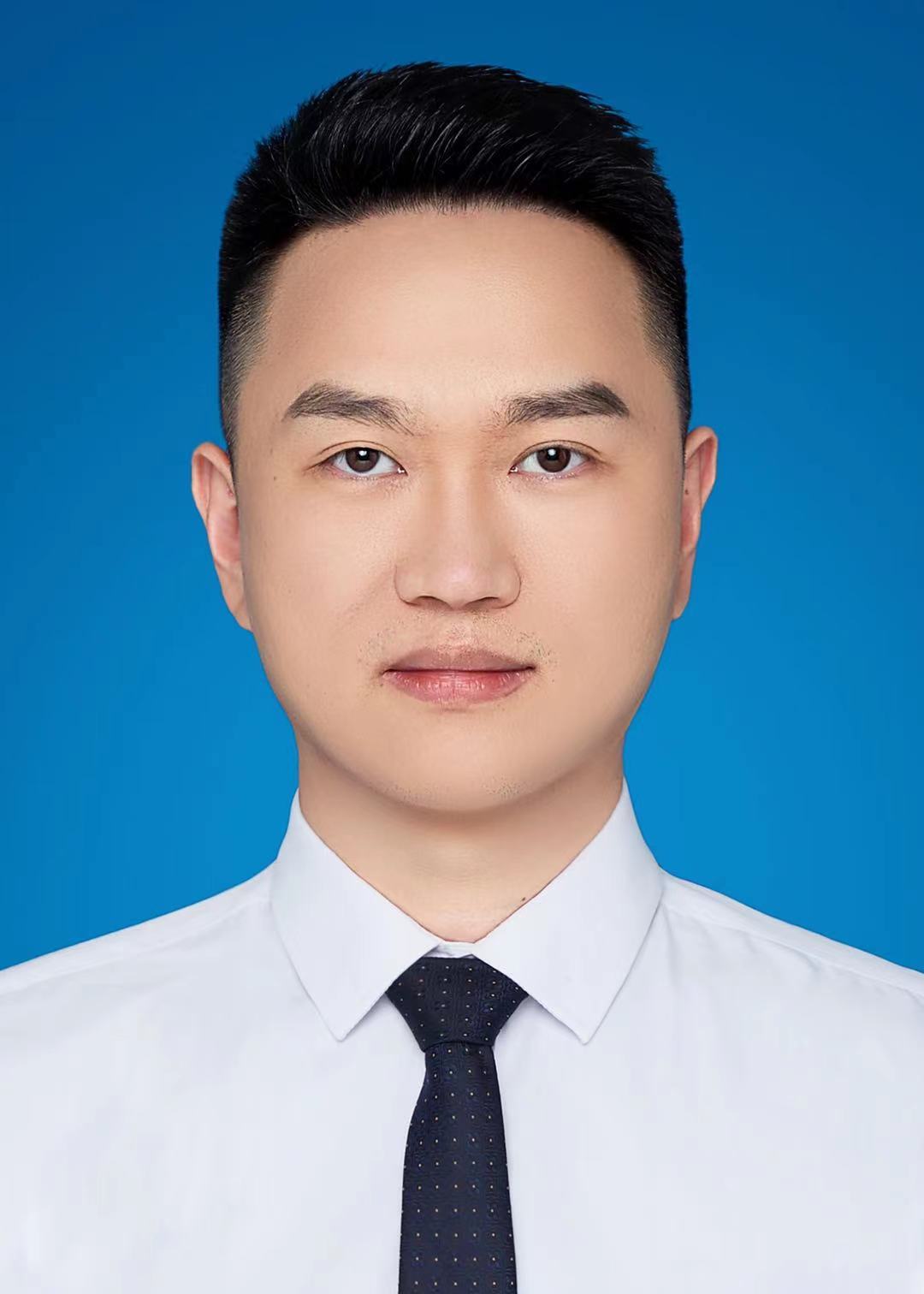}}]{Le Zhang} is a professor with the School of Information and Communication Engineering at the University of Electronic Science and Technology of China (UESTC). He earned his Ph.D. degree from Nanyang Technological University (NTU) in 2016. Following his graduation, he served as a postdoctoral fellow at the Advanced Digital Sciences Center (ADSC) in Singapore from 2016 to 2018. His career continued to evolve as he took on the role of a research scientist at the Institute for Infocomm Research (I2R) under the Agency for Science, Technology, and Research (A*STAR), Singapore, where he worked from 2018 to 2021. His research primarily focuses on computer vision and machine learning. He is an associate editor of Neural Networks, Neurocomputing, and IET Biometrics. Additionally, he has been a guest editor for several journals, including IEEE Transactions on Neural Networks and Learning Systems, IEEE Transactions on Big Data, Pattern Recognition and so on. He has been honored with multiple paper awards, including the 2022 Norbert Wiener Review Award in IEEE/CAA J. Autom. Sinica, as well as the Best Paper Awards at the 2022 IEEE HPCC and IEEE ICIEA conferences.
\end{IEEEbiography}

\begin{IEEEbiography}[{\includegraphics[width=1in,height=1.25in,clip,keepaspectratio]{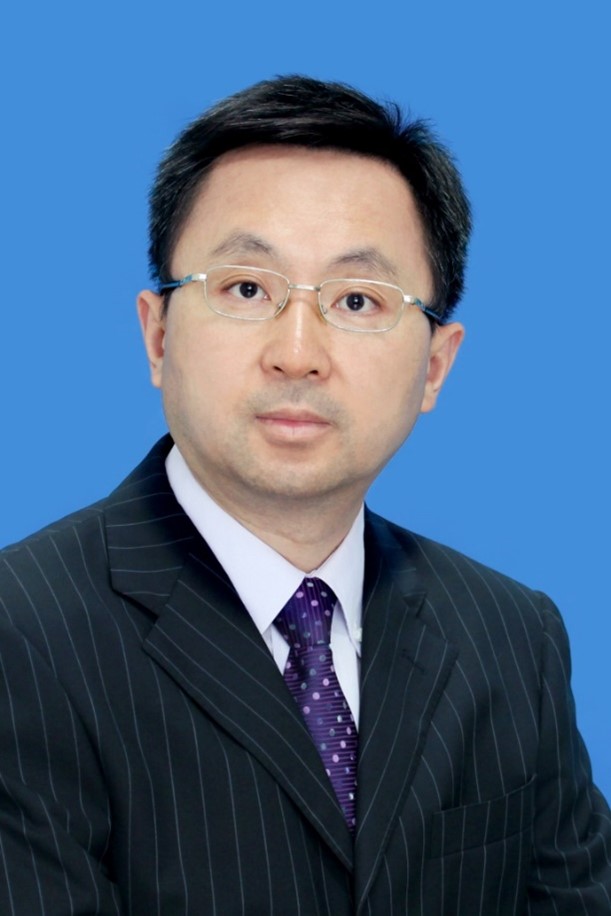}}]{Ce Zhu} (M’03–SM’04–F’17) received the B.S. degree from Sichuan University, Chengdu, China, in 1989, and the M.Eng and Ph.D. degrees from Southeast University, Nanjing, China, in 1992 and 1994, respectively, all in electronic and information engineering. He held a post-doctoral research position with the Chinese University of Hong Kong in 1995, the City University of Hong Kong, and the University of Melbourne, Australia, from 1996 to 1998. He was with Nanyang Technological University, Singapore, for 14 years from 1998 to 2012, where he was a Research Fellow, a Program Manager, an Assistant Professor, and then promoted to an Associate Professor in 2005. He has been with University of Electronic Science and Technology of China (UESTC), Chengdu, China, as a Professor since 2012, and serves as the Dean of Glasgow College, a joint school between the University of Glasgow, UK and UESTC, China. His research interests include video coding and communications, video analysis and processing, 3D video, visual perception and applications. He has served on the editorial boards of a few journals, including as an Associate Editor of IEEE TRANSACTIONS ON IMAGE PROCESSING, IEEE TRANSACTIONS ON CIRCUITS AND SYSTEMS FOR VIDEO TECHNOLOGY, IEEE TRANSACTIONS ON BROADCASTING, IEEE SIGNAL PROCESSING LETTERS, an Editor of IEEE COMMUNICATIONS SURVEYS AND TUTORIALS, and an Area Editor of SIGNAL PROCESSING: IMAGE COMMUNICATION. He has also served as a Guest Editor of a few special issues in international journals, including as a Guest Editor in the IEEE JOURNAL OF SELECTED TOPICS IN SIGNAL PROCESSING. He was an APSIPA Distinguished Lecturer (2021-2022), and also an IEEE Distinguished Lecturer of Circuits and Systems Society (2019-2020). He is serving as the Chair of IEEE ICME Steering Committee (2024-2025), and the Chair of IEEE Chengdu Section. He is a co-recipient of multiple paper awards at international conferences, including the most recent Best Demo Award in IEEE MMSP 2022, and the Best Paper Runner Up Award in IEEE ICME 2020.
\end{IEEEbiography}

\end{document}